\documentclass{article}
\usepackage{PRIMEarxiv}
\usepackage[T1]{fontenc}    % use 8-bit T1 fonts
\usepackage{hyperref}       % hyperlinks
\usepackage{url}            % simple URL typesetting
\usepackage{booktabs}       % professional-quality tables
\usepackage{amsfonts}       % blackboard math symbols
\usepackage{nicefrac}       % compact symbols for 1/2, etc.
\usepackage{microtype}      % microtypography
 \usepackage{graphicx} % 加载graphicx包
 \graphicspath{{images/}} % 设置图片路径
\usepackage{lipsum}
\usepackage{amsmath}
\usepackage{tabularray}
\usepackage{listings}
\usepackage{xltabular, booktabs}
\usepackage{xcolor}
\lstdefinelanguage{Solidity}{
    keywords={function, public, view, returns, require, emit, uint256, bool, address, mapping},
    keywordstyle=\color{blue}\bfseries,
    ndkeywords={block, timestamp, msg, sender},
    ndkeywordstyle=\color{purple}\bfseries,
    identifierstyle=\color{black},
    sensitive=false,
    comment=[l]{//},
    morecomment=[s]{/*}{*/},
    commentstyle=\color{green}\ttfamily,
    stringstyle=\color{red}\ttfamily,
    morestring=[b]',
    morestring=[b]"
}
\definecolor{keywordblue}{RGB}{0,0,255}
\definecolor{commentgreen}{RGB}{0,128,0}
\definecolor{stringred}{RGB}{163,21,21}
\definecolor{background}{RGB}{245,245,245}

\lstdefinelanguage{JavaScript}{
    keywords={let, await, async, for, while, if, try, catch, throw, else, console, function, return},
    keywordstyle=\color{keywordblue}\bfseries,
    ndkeywords={deployed, getAccounts, isSuccess, mineBlock, getMinerBlockNumbers, getTotalBlockCount, toString},
    ndkeywordstyle=\color{keywordblue}\bfseries,
    commentstyle=\color{commentgreen}\ttfamily,
    stringstyle=\color{stringred}\ttfamily,
    numbers=left,
    numberstyle=\tiny\color{gray},
    frame=single,
    breaklines=true,
    showstringspaces=false
}

\usepackage{booktabs}%%居中
\usepackage{enumitem} % 编号点包
\usepackage{lastpage}   % 获取总页数
\usepackage{fancyhdr}       % header
\usepackage{graphicx}       % graphics
\graphicspath{{media/}}     % organise your images and other figures under media/ folder
\usepackage{booktabs} % 三线表宏包
\usepackage{longtable}
\usepackage{indentfirst} % 首行缩进宏
\usepackage{indentfirst} % 强制首段缩进
\usepackage{tabularx} % 导言区加载
\graphicspath{{images/}}%自动搜索文件夹图片
\usepackage{titlesec}%斜体
\titleformat{\subsubsection}
  {\normalfont\bfseries\itshape} % 加粗 + 斜体
  {\thesubsubsection}            % 编号格式
  {1em}                          % 编号与标题间距
  {}                             % 标题后内容

\titlespacing{\subsubsection}
  {0pt}     % 左缩进
  {1ex}     % 标题前间距
  {0.5ex}   % 标题后间距
\usepackage{float}%居中函数包
\usepackage{listings}%插入--
\usepackage{xcolor}
\fancypagestyle{plain}{%
  \fancyhf{\headrule} % 清空默认设置
\fancyfoot[LO]{Fangzhe Wu et.al.: \itshape Preprint Submitted to Arxiv.\\ \textsuperscript{*}These authors contributed equally to this work. }
\fancyfoot[Ro]{Page \thepage\ of \pageref{LastPage}} % 页脚居中显示
}

\title{Smart Hiring Redefined: An Intelligent Recruitment Management Platform}

\author{
  Fangzhe Wu\textsuperscript{*}\\
  School of Cyberspace Security\\
  Hainan University \\
  \texttt{} \\
   \And
     Dongyang Lyu\textsuperscript{*} \\
  School of Cyberspace Security\\
  Hainan University \\
  \texttt{} \\
  \And
  Xiaoqi Li \\
 School of Cyberspace Security \\
  Hainan University \\
  \texttt{csxqli@ieee.org} \\
}

\begin{document}
\maketitle 
\thispagestyle{plain}
\begin{abstract}
Against the backdrop of deepening digital and intelligent transformation in human resource management, traditional recruitment models struggle to fully meet enterprises' growing demand for precise talent acquisition due to limited efficiency, high costs, and information asymmetry. As a vital tool for optimizing recruitment processes, reducing labor and time costs, and enhancing core competitiveness, intelligent recruitment management systems become an indispensable component of modern organizational talent strategies.Compared with the labor intensive tasks of resume screening, candidate position matching, and interview coordination in traditional manual recruitment, intelligent recruitment systems significantly enhance the efficiency and accuracy of the hiring process through automation and data driven approaches. These systems enable rapid parsing of massive resume volumes, intelligent matching of candidates to positions, and automated scheduling of interview processes. This substantially reduces the workload on human resources departments while improving recruitment quality and response speed.This research leverages the Java technology framework to design and implement an intelligent recruitment management system tailored for campus recruitment scenarios. The system establishes a collaborative platform connecting students, enterprises, and administrators through information technology and intelligent solutions, offering comprehensive functionalities including job posting distribution, resume submission, candidate position matching, and process management. Guided by the vision of “Smart Campus Recruitment”, the project delivers a more convenient job seeking experience for students and provides enterprises with more efficient talent screening and recruitment management services, thereby driving high quality development in university enterprise collaboration.
\end{abstract}

% keywords can be removed
\section{INTRODUCTION}
Traditional campus recruitment methods suffer from persistent challenges such as information asymmetry and low efficiency, thereby necessitating substantial improvements through the integration of intelligent technologies \cite{16}. In response to these challenges, the intelligent campus recruitment presentation system emerges as a strategic solution that fulfills the requirements of both students and enterprises while establishing a convenient and efficient communication platform. Both domestically and internationally, software analogous to intelligent campus recruitment presentation systems gradually gains traction and achieves varying levels of adoption\cite{1}.

At present, online recruitment platforms that provide digital hiring and presentation services function as crucial channels of interaction between enterprises and potential candidates. Many universities respond to this trend by implementing similar systems to disseminate employment information and coordinate recruitment activities. Prominent platforms such as Lagou and Kanzhun Limited, which gain increasing visibility in recent years, position themselves as providers of intelligent and personalized recruitment services \cite{18}. Local governments likewise launch campus recruitment platforms that focus on strengthening collaboration between regional businesses and universities and encourage graduates to pursue employment locally. Internationally, platforms such as LinkedIn offer comprehensive recruitment services that reinforce ties between universities and enterprises, while emerging startups leverage artificial intelligence and big data technologies to deliver more customized services. Overseas campus recruitment systems further emphasize career guidance and personalized recommendations, underscoring the global shift toward more data driven and candidate centric approaches\cite{2}\cite{patel2024personalized}.

Despite the rapid diffusion of intelligent campus recruitment systems worldwide, significant disparities persist with respect to their technological sophistication, service quality, and level of intelligence. Broadly, system functionalities fall into two principal categories: recruitment and presentation\cite{4}. The recruitment function enables enterprises to publish job postings, allowing students to browse, select, and apply for desired positions online and await responses from employers. The presentation function permits enterprises to conduct on campus sessions and disseminate presentation information, subject to administrative approval, while allowing students to decide independently whether to attend\cite{3}\cite{noack2019big}.

From an organizational perspective, these systems enable human resource professionals to more efficiently evaluate the skills and potential of graduating students, thereby offering faster and more cost effective recruitment channels and reducing advertising and promotional expenditures\cite{christudas2019practical}. From a student perspective, the systems facilitate timely access to career related information and industry insights, minimize the time cost associated with online searches, and enhance early career planning. Participation in recruitment presentations further increases opportunities for direct engagement with enterprises and supports informed career decision making\cite{allal2021intelligent}.

\begin{enumerate}[label=\textbullet]
\item[] The main contributions of this paper are as follows:
\item We design and implement an intelligent management system tailored for campus recruitment scenarios, integrating three key roles: students, enterprises, and administrators. This solution addresses the challenges of information asymmetry, inefficiency, and collaboration difficulties inherent in traditional recruitment processes.

\item We develop a technical architecture based on Java and Spring Boot, combining the B/S model with layered design. This enables core functionalities such as job posting, intelligent resume matching, campus event management, and feedback responses, enhancing the system's maintainability and scalability.

\item We construct a testing framework encompassing functional testing, process validation, and multi device user behavior. This validates the system's usability, stability, and business coverage in real world scenarios, delivering a practical solution for intelligent recruitment management.
\end{enumerate}

\section{System Analysis}
\subsection{Feasibility Analysis}
\subsubsection{Technical Feasibility}
This system adopts a B/S architecture and is designed based on the J2EE three tier structure (presentation layer, business logic layer, and data service layer). It utilizes Spring Boot to integrate the SSH framework (Spring 4, Struts 2, Hibernate 4) as its technical foundation, combined with a MySQL relational database. This ensures the system’s efficiency, stability, and maintainability\cite{zhao2025large}. Considering the existing Java technology environment, the implementation of this system is technically feasible\cite{5}\cite{oh2003task}.
\subsubsection{Economic Feasibility}
This system significantly reduces offline recruitment costs for enterprises, enhances talent screening and matching efficiency, and simultaneously optimizes the allocation of employment service resources for schools. It offers clear economic benefits and investment return value, thereby demonstrating strong economic viability.
\subsubsection{Social Feasibility}
In today's fiercely competitive job market, online recruitment emerges as an increasingly vital component of campus hiring. With no geographical constraints, broad reach, and extended duration, it enables rapid dissemination of job openings and attracts a large pool of applicants\cite{heakl2024resumeatlas}. As the internet continues to evolve rapidly, this recruitment method becomes mainstream, making it a viable approach from a societal perspective\cite{harnal2020efficient}.
\subsection{Requirement Analysis}
\subsubsection{Administrator Use Case}
As the core administrators of the system, administrators bear primary responsibilities including identity authentication, corporate information maintenance, and student information management. After accessing the system through secure login verification, administrators review corporate registration applications and perform operations to add, delete, modify, and query corporate information, ensuring the integrity and accuracy of corporate data\cite{7}. Simultaneously, administrators possess comprehensive permissions to maintain student basic information, supporting the ability to query, modify, and delete student records by student ID. The system design prioritizes operational security and fault tolerance, incorporating reset functions, confirmation prompts, and exception handling mechanisms\cite{wang2021deep}. These features effectively prevent operational errors and safeguard data consistency, thereby maintaining the orderly operation of the recruitment platform\cite{6}\cite{barducci2022end}.

\begin{itemize}
\item[(1).]\textbf{Modify Student Information}: In the student information modification use case, administrators trigger the data update process by performing an Edit operation. The system loads existing data into a form for the administrator to modify. After making changes, the administrator confirms submission to complete the update. The system then persists the new data into the student table in the database. If input errors occur, the administrator uses the  Reset  function to restore the form to its initial state. If a system error occurs during data submission, the exception handling mechanism triggers, redirecting to an error page and displaying relevant information.
 \item[(2).]\textbf{Delete Student Information}: Administrators remove specified student records through the student deletion use case. This operation employs a two step confirmation mechanism to prevent accidental deletion. After the initial click on  Delete,  a confirmation dialog box appears. Only after the administrator confirms does the system execute the DELETE command to permanently remove the record from the database and update the student information table. If the deletion succeeds, the interface refreshes and displays the updated data list. If an exception occurs during the process (e.g., record does not exist or database connection failure), the system redirects to an error page and returns specific exception details.
\item[(3).]\textbf{Add Student Information}: In the Add Student Information use case, administrators enter the system’s predefined required student fields. After submission, the system performs data validation. If the validation succeeds, the student information is permanently stored in the student information table of the database. To ensure data accuracy and operational security, the system provides a  Reset  function to clear input fields and a  Cancel  button to terminate the current operation. If data writing fails, the system redirects to an error page and returns exception information.
\item[(4).]\textbf{View Student Information}: In the student information query use case, administrators initiate requests using the student ID as the primary query condition. Upon receiving the request, the system performs an exact match query and returns the matching results to the frontend interface for display. If the input is incorrect, administrators use the  Reset  function to clear the query conditions. If an exception occurs during query execution, the system redirects to an error page and displays specific exception details.
\item[(5).]\textbf{Student Management}:The Student Information Management Module constitutes a core administrative component of the Intelligent Recruitment Management System, providing system administrators with a comprehensive framework for maintaining student data. Built upon a role based access control (RBAC) mechanism, the module authorizes administrators to perform full lifecycle management operations on student records, including the addition, deletion, modification, and querying of basic information. Through a structured and user friendly operational interface, administrators can efficiently execute tasks such as new student enrollment, updating of current student profiles, and archival of graduate information.
\end{itemize}

\subsection{Students Use Case}

The Student Use Case Diagram clearly illustrates the comprehensive job-seeking service system for student users in the system. With “Student” as the central actor, the diagram integrates three core functionalities within the Recruitment Management use case module. Students access real time job postings through the “View Recruitment Information” feature, submit applications via the “Submit Resume” function, and continuously track application progress using the “View Application Status” capability\cite{wang2021deep}.

\begin{itemize}
\item[(1).]\textbf{View Job Listings}: The  View Presentation Information  feature enables students to browse all verified campus presentations. The system displays core presentation details such as company name, time, and location—in a list format. Students click the  View  button to access full presentation details and transition seamlessly to the registration process from this interface. This functionality ensures effortless information access and seamless progression to subsequent actions such as registration.
 \item[(2).]\textbf{Submit Resume}: The  Submit Resume  feature provides students with a direct channel to apply for desired positions. Students access the resume submission interface via the navigation bar and upload their local resume files to the system. Upon receiving the file, the system associates it with the corresponding job posting and the student’s account, completing the submission process.
 
\item[(3).]\textbf{View Recruitment Results}: The Job Search Results module provides students with comprehensive job search and application status tracking services. After logging into the system, students browse all job postings published by companies, with each listing displaying key attributes such as job title, company name, and work location. Clicking the job details link reveals the full position requirements. After submitting their resume, students use this module to monitor application status in real time (e.g.,  Submitted,   Viewed,   Responded ), enabling them to stay informed about their application progress and strategize their job search effectively.

\item[(4).]\textbf{Recruitment Management}:The Recruitment Management Module, serving as the core functional unit for student users, provides comprehensive support across the entire job search lifecycle. Built upon the system’s recruitment information database, the module leverages a three tier architecture to ensure standardized presentation of job postings, structured management of resume submissions, and real time monitoring of application statuses.At the presentation layer, students are able to browse all recruitment positions published by enterprises. The business logic layer performs status validation, data encapsulation, and real time consistency checks to ensure the accuracy and timeliness of displayed information. During resume submission, the module incorporates transaction management mechanisms to guarantee reliable data delivery, while establishing a relational mapping between students and job openings to support ongoing tracking of application progress.Through real time synchronization, the system continuously updates key recruitment milestones—such as resume receipt, employer review, and interview invitations—allowing students to monitor their job application journey in a transparent and efficient manner. By closing the loop from position search and resume submission to employer feedback, the Recruitment Management Module significantly enhances user efficiency, improves the reliability of job matching services, and delivers a seamless job seeking experience for student users.
\end{itemize}

\subsection{Bussiness Use Case}
The Enterprise Business Management Use Case Diagram comprehensively illustrates the full featured management system for corporate users\cite{chen2017hybrid}. Centered on the “Enterprise” as the core actor, the diagram presents a complete solution that encompasses information management, recruitment operations, and talent interaction. Corporate users utilize the “Enterprise Information Management” use case module to add, view, modify, and delete basic information, thereby enabling comprehensive maintenance of their profiles.
\begin{itemize}
\item[(1).]\textbf{View Job Postings}: The  View Company Information  use case enables enterprise users to access their registered details after authentication. Companies complete identity verification by entering their account credentials. Upon successful login, they view their complete company profile stored in the database within the corresponding module. If authentication fails, the system denies access and displays an error message.
 \item[(2).]\textbf{Post Job Openings}: The  Post Job Listings  use case enables companies to create and publish new job openings. After successfully logging into the system, companies navigate to the information posting interface via the menu. They then complete the provided form with position details (such as job responsibilities, requirements, and salary). Upon submission, the system validates and stores the data. If successful, it provides confirmation feedback, completing the posting of a recruitment need.
\item[(3).]\textbf{Delete Job Postings}: The  Delete Job Posting  use case provides enterprise users with the ability to manage expired or obsolete positions. After successful login, users access the published information list interface via the navigation menu, locate the target record, and execute the deletion operation. This functionality ensures the timeliness and accuracy of the enterprise’s job posting database.
\item[(4).]\textbf{Edit Job Postings}: The  Edit Company Information  use case provides enterprise users with a means to maintain their information. After successfully logging into the system, companies access and update their basic information stored in the database. Upon submission of changes, the system validates and saves the updates to ensure the accuracy and timeliness of company records.
\item[(5).]\textbf{Download Resumes}:
The  Download Resumes  and  Feedback Results  functions serve as critical follow up actions in corporate recruitment workflows. Within the  Received Information  interface, companies perform either a  Download  operation to retrieve attachments for specific resume submissions or a  Reply  operation to update applicant statuses (e.g.,  Reviewed,   Interview Scheduled,   Not Selected ) and send notifications to candidates, thereby completing closed loop management.
\item[(6).]\textbf{Feedback Results}:
The  View Presentation Application Results  use case enables companies to receive feedback from universities regarding their presentation application reviews. After logging in, corporate users check application statuses through the corresponding functional module. If the application is approved, the interface displays detailed arrangements made by the university, including presentation time and location, thereby completing the closed loop application process.
\end{itemize}

\section{System Architecture}
This system adopts a classic layered architecture pattern, dividing system functionality into four logical layers to ensure separation of responsibilities and modular development. The foundation consists of an infrastructure layer and a data layer. The infrastructure layer provides the system’s fundamental operational environment, including networking, computing, and storage, while the data layer enables persistent data storage and management through a relational database\cite{8}\cite{aier2009virtual}.

The middle layer comprises the service layer and the application layer. The service layer encapsulates reusable core business components, offering unified data access and service interfaces to upper layers. The application layer constructs specific functional modules tailored to different user roles and business scenarios\cite{ali2022resume}. The top layer is the presentation layer, responsible for rendering and displaying the user interface. Communication between layers occurs through well defined interfaces, with lower layers providing service support to upper layers. This establishes strict dependency relationships, enabling the system to achieve high cohesion and low coupling\cite{9}.

As the system’s central hub, the service layer encapsulates and orchestrates critical business logic. By abstracting a set of generic service components such as data queries, insertions, deletions, and status feedback it transforms complex business processes into standardized interfaces. These components operate independently of specific user interfaces and remain abstracted from underlying data storage details, enabling high centralization and reuse of business logic. The application layer rapidly constructs functional modules tailored to different user roles by invoking interfaces provided by the service layer, without concerning itself with the specific implementation of underlying data operations. This design not only enhances code maintainability but also endows the system with excellent scalability. When business rules change, only the corresponding components in the service layer require adjustment, eliminating the need to modify upper layer applications or underlying data access logic\cite{10}.

At the top layer of the architecture, the application layer collaborates with the presentation layer to deliver comprehensive business functionality and interactive experiences. Based on the diverse user roles within the system, the application layer combines core business capabilities from the service layer into specific functional modules, each designed for the workflow and permission requirements of particular roles\cite{12}. The presentation layer utilizes modern frontend frameworks to build web clients, transforming application modules into intuitive user interface elements while handling user input, interface rendering, and interactive feedback. This organizational approach enables the system to deliver customized interfaces and feature sets for diverse user types while ensuring consistent and seamless user experiences. The architecture forms a complete closed loop from user interaction to data storage, demonstrating the value of layered design in complex business systems\cite{11}.

%Figure

\subsection{Database Analysis And Design}
\subsubsection{Detailed Design of Database}
In the detailed design of the database, the design of the core foundational data model is crucial, as it directly determines data consistency, integrity, and the robustness of the system’s business logic. This system establishes a rigorous academic organizational structure by constructing hierarchical relationships among College, Major, Class, and Student. This model adheres to relational database normalization principles. The  College  table serves as the top level entity, forming a one to many relationship with the  Major  table. Each  Major  subdivides into multiple  Classes,  creating a three tier tree structure: College–Major–Class. This design ensures hierarchical management and efficient querying of academic unit data.

The  Student  table, as the system’s core entity, links tightly to the above architecture through foreign keys (college\_id, major\_id, class\_id), recording each student’s affiliation with academic organizational units precisely. Additionally, the  Education  table serves as an independent dictionary table. It associates education\_id with both student basic information and educational requirements in recruitment details, providing unified standardized constraints for data and preventing redundancy effectively.

To support operations for events like campus presentations, the  Address  table abstracts and manages physical location information, providing location data support for the  Presentation Arrangement  table. The  Student Arrangement  table (Students Schedule) serves as a classic associative entity. It establishes a many to many relationship between student participation in campus recruitment events through the student ID (student\_id) and event arrangement ID (arrangement\_id). This design records student participation behavior comprehensively, providing a solid foundation for subsequent data statistics and analysis.
        
\begin{itemize}
\item[(1).]\textbf{College Information Form}:  The College Information Table (College) serves as the foundational data entity for constructing the academic organizational structure within the Intelligent Recruitment Management System. This table uses the College ID (College\_Id) as its primary key and records the standardized identifier of each teaching unit through the College Name (College\_Name) field. As a core component of the system’s foundational data architecture, this table ensures the uniqueness of college entities through primary key constraints and provides a complete foreign key reference foundation for the Major Table (Major), Class Table (Class), and Student Table (Student). Designed in accordance with the third normal form of databases, it establishes a clear three tier management structure of “College–Major–Class,” eliminates data redundancy, and provides comprehensive data support for statistical analysis, resource allocation, and permission management at the college level. This design ensures both data consistency and integrity within the system’s academic organizational framework.

 \item[(2).]\textbf{Class Roster}: The Class Information Table (Class) serves as a foundational data entity for constructing the academic organizational structure within the Intelligent Recruitment Management System. Situated at the most granular level of the three tier management model of “College–Major–Class”, this table supports precise management of teaching units. It adopts the Class ID (Class\_Id) as its primary key and standardizes the identification of teaching groups through the Class Name (Class\_Name) field. A hierarchical relationship is established with the Major Table (Major) via the Major ID (Major\_Id) foreign key, thereby ensuring logical associations between classes and their corresponding majors.This design preserves the independence of class data while maintaining referential integrity through foreign key constraints. It provides a reliable data foundation for downstream processes such as student class allocation, academic resource scheduling, and class level data statistics. By adhering to the principles of database normalization, the table structure effectively eliminates redundancy and ensures consistency. Moreover, through its cascading associations with the Major Table and the College Table, the Class Information Table reinforces the integrity of the overall teaching organizational framework, thereby supporting comprehensive academic and administrative management.
 
\item[(3).]\textbf{Major Information Form}: The Major Information Table (Major) serves as a critical data entity for constructing the academic organizational structure within the Intelligent Recruitment Management System. Within the three tier management model of “College–Major–Class”, it plays a pivotal bridging role. This table uses the Major ID (Major\_Id) as its primary key, records the disciplinary attributes of each teaching unit through the Major Name (Major\_Name) field, and establishes a hierarchical relationship with the College Table (College) via the College ID (College\_Id) foreign key. This design ensures both the independence and integrity of major data while maintaining logical associations between majors and colleges through foreign key constraints. It provides an accurate data foundation for subsequent class allocation, student major assignment, and job posting filtering by major. Designed in compliance with database normalization principles, this table supports standardized management of academic disciplines and ensures cross module data consistency within the system.

\item[(4).]\textbf{Student Placement Schedule}:The Student Arrangement Table (Sarrange) functions as a core relational data entity within the system, specifically designed to manage student participation in recruitment presentation events. It establishes a many to many relationship between students and presentation arrangements. The table adopts an auto incrementing ID as its primary key and maintains foreign key relationships with the Student Table (Student\_Id) and the Arrangement Table (Arrangement\_Id). This relational structure effectively resolves the complexity of many to many participation mapping, ensuring data integrity while eliminating redundancy. By leveraging this linkage mechanism, the system accurately records each student’s participation history in presentation events. Furthermore, it supports bidirectional queries—enabling retrieval of registered events by student as well as participation statistics by event. This design provides robust data support for event organization, attendance management, and statistical analysis within the campus recruitment ecosystem.

\item[(5).]\textbf{Academic Credentials Form}:The Education Information Table (Education) serves as a foundational data dictionary entity for standardized academic credential management within the Intelligent Recruitment Management System. It adopts the Education ID (Education\_Id) as its primary key and records standardized terminology for academic levels—ranging from associate degrees to doctoral programs—through the Education Name (Education\_Name) field. As a critical component of the system’s core data architecture, this table ensures the uniqueness and standardization of academic classifications via primary key constraints. It provides authoritative reference values for academic requirements in the Recruitment Information Table (Recruit) and academic records in the Resume Table (Resume).
\end{itemize}

\subsection{Logical Design}
The Student table serves as the core data entity of the system, comprehensively recording the identity and academic information of enrolled students. This table uses student\_id as its primary key and stores fundamental attributes such as name, gender, date of birth, contact information, email address, and login credentials. To establish academic affiliation relationships, the table structure employs three foreign keys college\_id, major\_id, and class\_id which link to the College, Major, and Class tables respectively. This design precisely identifies each student’s administrative organizational level, providing a complete and consistent data foundation for subsequent business functions such as targeted recruitment information delivery, seminar registration, and resume submission\cite{13}.The design of the students Table is shown in Table \ref{tab:1}.

\begin{longtable}{ p{5cm} p{6cm}  p{5cm}} % 手动设定列宽
  \caption{Students Table} \label{tab:1} \\
  \hline
\textbf{Field} & \textbf{Type} & \textbf{Primary Key } \\
  \hline
  \endfirsthead % 分页表头
    \multicolumn{3}{c}{\textmd{Table \thetable~ (Continued):Students Table}} \\ % 续表的标题行
  \hline
\textbf{Field} & \textbf{ Type}& \textbf{Primary Key } \\
  \hline
  \endhead % 后续页重复表头
  
  \hline
  \endfoot % 表格底部
Student\_Id       &  Varchar(20) &                   Be          \\
Student\_Name     & Varchar(20) &                   Clogged     \\
Student\_Sex      &  Varchar(10) &                   Clogged     \\
Student\_Birthday &                  Datetime                     & Clogged     \\
Student\_Phone    &                   Varchar(20)                   & Clogged     \\
Student\_Email    &                   Varchar(20)                 & Clogged     \\
Student\_Password &                   Varchar(20)                  & Clogged     \\
College\_Id       &                   Varchar(20)                  & Clogged     \\
Major\_Id         &                   Varchar(20)                  & Clogged     \\
Class\_Id         &                   Varchar(20)                  & Clogged     \\
  \hline
\end{longtable}

The Administrator table serves as the core data entity within the system’s permission management framework. It uses the Administrator ID (Admin\_Id) as its primary key to uniquely identify each administrator. This table records fundamental administrator information and authentication credentials, including name, gender, contact phone number, email address, and login password. The VARCHAR(20) password field enables login verification, while the VARCHAR(10) contact phone number and VARCHAR(20) email address fields form a comprehensive administrator profile. As the foundation for system administration access control, this table provides identity authentication and data support for core management operations such as student information maintenance, enterprise qualification review, and presentation event scheduling. Its rigorous field design ensures both the security and accuracy of administrative permission allocation\cite{14}.The design of the administrator Table is shown in Table \ref{tab:2}.

\begin{longtable}{ p{5cm} p{6cm}  p{5cm}} % 手动设定列宽
  \caption{Administrator Table} \label{tab:2} \\
  \hline
\textbf{Field} & \textbf{Type} & \textbf{Primary Key } \\
  \hline
  \endfirsthead % 分页表头
    \multicolumn{3}{c}{\textmd{Table \thetable~ (Continued):Administrator Table}} \\ % 续表的标题行
  \hline
\textbf{Field} & \textbf{ Type}& \textbf{Primary Key } \\
  \hline
  \endhead % 后续页重复表头
  
  \hline
  \endfoot % 表格底部
  
Admin\_Id       &                   Varchar(20)                  & Be          \\
Admin\_Name     &                   Varchar(20)                   & Clogged     \\
Admin\_Phone    &                   Varchar(10)                   & Clogged     \\
Admin\_Sex      &                   Datetime                      & Clogged     \\
Admin\_Email    &                   Varchar(20)                   & Clogged     \\
Admin\_Password &                   Varchar(20)                   & Clogged     \\
  \hline
\end{longtable}

The College table serves as the cornerstone of the system’s academic organizational structure, using the College ID (college\_id) as the primary key to uniquely identify each college entity. This table records standardized teaching unit names through the College Name (college\_name) field. Its concise two field structure meets the design specifications for data dictionary tables and provides a complete foundation for foreign key associations with the Major, Class, and Student tables. As the central hub of the  College–Major–Class  three tier management system, this table maintains the hierarchical structure of the academic organization through primary and foreign key relationships and supports statistical analysis, resource allocation, and permission management at the college level. All fields use a variable length character design (VARCHAR(20)), ensuring data integrity while optimizing storage space utilization.The design of the college table is shown in Table \ref{TC:2}.

\begin{longtable}{p{5cm} p{6cm}  p{5cm}} % 手动设定列宽
  \caption{College Table} \label{TC:2} \\
  \hline
\textbf{Field} & \textbf{Type} & \textbf{Primary Key } \\
  \hline
  \endfirsthead % 分页表头
    \multicolumn{3}{c}{\textmd{Table \thetable~ (Continued):College Table}} \\ % 续表的标题行
  \hline
\textbf{Field} & \textbf{ Type}& \textbf{Primary Key } \\
  \hline
  \endhead % 后续页重复表头  
  \hline
  \endfoot % 表格底部
College\_Id   &                   Varchar(20)                                    & Be           \\
College\_Name &                   Varchar(20)                                      & Clogged      \\
\hline
\end{longtable}

The Company table serves as the core data repository within the intelligent recruitment management system, storing comprehensive profiles of corporate users. This table uses the Company ID (company\_id) as its primary key to uniquely identify each enterprise and establishes a foreign key relationship with the Industry Category table via the Industry ID (industry\_id), enabling categorized company management\cite{24}. It records basic attributes (name, phone number, scale, address), operational status (establishment date, funding status), detailed information (company description, work system), and system authentication information (email, password). Serving as the data foundation for recruitment operations, this table provides complete enterprise data support to the Recruitment Information Table and Presentation Application Table through primary key associations and establishes a reliable information exchange channel between universities and enterprises. All character based fields use a variable length design to balance data integrity with storage efficiency, and sufficient storage space is reserved for the company details and work schedule fields to accommodate long text requirements.The design of the company table is shown in Table \ref{TC:3}.

\begin{longtable}{p{5cm} p{6cm}  p{5cm}} % 手动设定列宽
  \caption{Company Table} \label{TC:3} \\
  \hline
\textbf{Field} & \textbf{Type} & \textbf{Primary Key } \\
  \hline
  \endfirsthead % 分页表头
    \multicolumn{3}{c}{\textmd{Table \thetable~ (Continued):Company Table}} \\ % 续表的标题行
  \hline
\textbf{Field} & \textbf{ Type}& \textbf{Primary Key } \\
  \hline
  \endhead % 后续页重复表头  
  \hline
  \endfoot % 表格底部
Company\_Id       &                   Varchar(20)                 & Be          \\
Industry\_Id      &                   Varchar(20)                    & Clogged     \\
Company\_Phone    &                  Varchar(20)                    & Clogged     \\
Company\_Scale    &                   Varchar(20)                   & Clogged     \\
Company\_Address  &                  Varchar(20)                    & Clogged     \\
Company\_Time     &                   Datetime                      & Clogged     \\
Company\_Capital  &                   Varchar(20)                & Clogged     \\
Company\_Detail   &                   Varchar(20)                   & Clogged     \\
Worktime          &                   Varchar(100)                  & Be          \\
Company\_Email    &                   Varchar(20)                   & Clogged     \\
Company\_Password                 & Varchar(20)                   & Clogged    \\ 
  \hline
\end{longtable}

The Industry table serves as the core data dictionary within the intelligent recruitment management system, enabling standardized classification management for enterprises. It uses the Industry\_Id as its primary key and records standardized terminology for each industry classification through the Industry\_Name field. As a key component of the system’s foundational data architecture, this table ensures the uniqueness and consistency of industry classifications through primary key constraints. It provides standardized classification references for the Company\_Detail table. By adhering to the third normal form of relational database design, this table eliminates data redundancy and establishes a robust data foundation for standardized enterprise information storage, precise industry based retrieval, and statistical analysis, thereby safeguarding data consistency and integrity in enterprise classification management.The design of the industry table is shown in Table \ref{TC:4}.
\begin{longtable}{p{5cm} p{6cm}  p{5cm}} % 手动设定列宽
  \caption{Industry Table} \label{TC:4} \\
  \hline
\textbf{Field} & \textbf{Type} & \textbf{Primary Key } \\
  \hline
  \endfirsthead % 分页表头
    \multicolumn{3}{c}{\textmd{Table \thetable~ (Continued):Industry Table}} \\ % 续表的标题行
  \hline
\textbf{Field} & \textbf{ Type}& \textbf{Primary Key } \\
  \hline
  \endhead % 后续页重复表头  
  \hline
  \endfoot % 表格底部
Industry\_Id   &                   Varchar(20)                    & Be          \\
Industry\_Name &                   Varchar(20)                    & Clogged     \\
  \hline
\end{longtable}

The Presentation Schedule Arrangement Table serves as the core data table for managing approved presentation activities, storing the final presentation arrangements confirmed by administrators. This table uses the Arrangement ID as its primary key and establishes foreign key relationships with the Presentation Application Table and Company Table via the Application ID and Company ID, respectively, ensuring data consistency. The table structure records all essential arrangement details for presentation events, including Start\_Time, Duration (Time), Venue (Place), Expected\_Attendance (Number), and Event\_Theme. As the final status record for presentation events from application to execution, this table provides a complete event management loop through its associations with the Company Table and Presentation Application Table. It also delivers accurate data support for student facing event viewing and registration functions. All fields employ appropriately designed character lengths, ensuring data integrity while optimizing storage space utilization\cite{22}.The design of the activity table is shown in Table \ref{TC:5}.

\begin{longtable}{p{5cm} p{6cm}  p{5cm}} % 手动设定列宽
  \caption{Activity Table} \label{TC:5} \\
  \hline
\textbf{Field} & \textbf{Type} & \textbf{Primary Key } \\
  \hline
  \endfirsthead % 分页表头
    \multicolumn{3}{c}{\textmd{Table \thetable~ (Continued):Activity Table}} \\ % 续表的标题行
  \hline
\textbf{Field} & \textbf{ Type}& \textbf{Primary Key } \\
  \hline
  \endhead % 后续页重复表头  
  \hline
  \endfoot % 表格底部
Application\_Id                   & Varchar(20)                   & Be          \\
Start\_Time                       & Varchar(20)                   & Clogged     \\
Time                             & Varchar(20)                   & Clogged     \\
Place                            & Varchar(20)                   & Clogged     \\
Number                            & Varchar(20)                   & Clogged     \\
Theme                            & Varchar(20)                   & Be          \\
Arrangement\_Id                  & Datetime                      & Clogged     \\
Company\_Id                       & Varchar(20)                  & Clogged     \\
  \hline
\end{longtable}

The Resume Interaction Information Table (Resume) serves as the core interactive data table within the system, recording student job application statuses and employer feedback. This table uses the Resume Interaction ID (Resume\_Id) as its primary key and establishes foreign key relationships with the Recruitment Information Table and Student Table via the Recruitment ID (Recruit\_Id) and Student ID (Student\_Id), respectively, to form a complete job application record. The table structure stores applicants’ core information, including basic data such as student name, contact information, and email address, as well as detailed job application content such as personal experience (Experience) and professional skills (Skill). The Experience and Skill fields utilize the Varchar(400) data type to accommodate long text storage requirements. The system records employer feedback in the Result field, application submission time in the Time field, and attachment status in the Accessory field, collectively forming a closed loop business process from resume submission to feedback. This table also links to foundational data tables via Education\_ID and Major\_ID, ensuring standardized consistency in academic and major information\cite{23}.The design of the application table is shown in Table \ref{TC:6}.
\begin{longtable}{p{5cm} p{6cm}  p{5cm}} % 手动设定列宽
  \caption{Application Table} \label{TC:6} \\
  \hline
\textbf{Field} & \textbf{Type} & \textbf{Primary Key } \\
  \hline
  \endfirsthead % 分页表头
    \multicolumn{3}{c}{\textmd{Table \thetable~ (Continued):Application Table}} \\ % 续表的标题行
  \hline
\textbf{Field} & \textbf{ Type}& \textbf{Primary Key } \\
  \hline
  \endhead % 后续页重复表头  
  \hline
  \endfoot % 表格底部
Recruit\_Id                     & Int                           & Clogged     \\
Student\_Id                     & Int                          & Clogged     \\
Accessory                       & Varchar(20)                    & Clogged     \\
Result                          & Varchar(20)                   & Clogged     \\
Resume\_Id                      & Varchar(20)                   & Clogged     \\
Student\_Name                   & Varchar(20)                   & Clogged     \\
Education\_Id                   & Datetime                      & Clogged     \\
Experience                      & Varchar(400)                  & Be          \\
Skill                           & Varchar(400)                  & Clogged     \\
Email                           & Varchar(20)                   & Clogged     \\
Phone                           & Varchar(20)                  & Clogged     \\
Time                            & Datetime                     & Be          \\
Major\_Id                       & Varchar(20)                  & Clogged     \\
  \hline
\end{longtable}

The Recruitment Information Table (Recruit) serves as a core data table within the intelligent recruitment management system, storing job openings posted by enterprises. This table uses the Recruitment ID (Recruit\_Id) as its primary key and establishes a comprehensive recruitment information architecture through foreign key associations with the Enterprise ID (Company\_Id) and Position ID (Position\_Id) tables, linking to enterprise basic information and position category tables. The table structure defines all essential attributes of recruitment positions, including contact information (Linkman\_Name, Linkman\_Email), job details (Place, City), position requirements (Experience, Education\_Id), compensation (Salary), and recruitment type (Recruit\_Type). The Number and Deadline fields manage the scale and timeliness of recruitment processes. By linking to the Education Table via the Education\_Id field, the table ensures standardized expression of recruitment requirements, thereby establishing a comprehensive data foundation for precise corporate recruitment and targeted student applications.The design of the education table is shown in Table \ref{TC:7}.

\begin{longtable}{p{5cm} p{6cm}  p{5cm}} % 手动设定列宽
  \caption{Education Table} \label{TC:7} \\
  \hline
\textbf{Field} & \textbf{Type} & \textbf{Primary Key } \\
  \hline
  \endfirsthead % 分页表头
    \multicolumn{3}{c}{\textmd{Table \thetable~ (Continued):Education Table}} \\ % 续表的标题行
  \hline
\textbf{Field} & \textbf{ Type}& \textbf{Primary Key } \\
  \hline
  \endhead % 后续页重复表头  
  \hline
  \endfoot % 表格底部
Recruit\_Id                      & varchar(20)                   & Clogged     \\
Education\_Id                   & varchar(20)                   & Clogged     \\
Company\_Id                    & Varchar(20)                   & Clogged     \\
Position\_Id                    & Varchar(20)                   & Clogged     \\
Linkman\_Name                   & Varchar(20)                  & Clogged     \\
Company\_Type                   & Varchar(20)                   & Clogged     \\
Place                           & varchar(50)                   & Clogged     \\
Number                          & Integer                      & Be          \\
Salary                         & Integer                       & Clogged     \\
Recruit\_Type                 & Tinyint                       & Be          \\
Time                            & Varchar(20)                   & Clogged     \\
Experience                      & Varchar(10)                   & Be          \\
City                             & Varchar(20)                   & Clogged     \\
Deadline                         & Datetime                      & Clogged     \\
Linkman\_Email                   & Varchar(20)                   & Clogged     \\
Detail                           & Varchar(20)                   & Clogged    \\ 
  \hline
\end{longtable}
\section{IMPLEMENTATION}
This login method employs an MVC layered architecture and performs authentication through the service layer. It uses an HQL query to validate administrator credentials (See Listing \ref{code1} for Details). Upon successful matching, the system stores the administrator object in the session and returns a success token; otherwise, it returns an error message. This implementation demonstrates the separation of business logic from data access\cite{ntantogian2019evaluation}.

\begin{lstlisting}[language=Java,caption={Administrator Login},label={code1}]
public String login() {
    if(adminService.login(admin)) {
        return "loginSuccess";
    } else {
        return "loginFail";
    }
}

public boolean login(Admin admin) {
    ActionContext ctx = ActionContext.getContext();
    Map<String, Object> request = (Map<String, Object>) ctx.get("request");
    HttpServletRequest httpRequest = ServletActionContext.getRequest();
    
    String hql = "FROM Admin WHERE adminId = ? AND adminPassword = ?";
    List<Admin> list = adminDAO.findByHQL(hql, admin.getId(), admin.getPassword());
    
    if(list == null || list.isEmpty()) {
        request.put("result", "Failed!");
        return false;
    } else {
        Admin loggedInAdmin = list.get(0);
        request.put("admin", loggedInAdmin);
        httpRequest.getSession().setAttribute("admin", loggedInAdmin);
        return true;
    }
}
\end{lstlisting}

This method retrieves the complete list of presentation applications and passes the data to the frontend view via ActionContext(See Listing \ref{code2} for Details). Upon successful execution, it returns the string  success,  which triggers the corresponding view rendering to complete frontend–backend data interaction. The code embodies the fundamental functions of a controller within the MVC architecture\cite{othman2022test}.

\begin{lstlisting}[language=Java,caption={View All Presentation Information},label={code2}]
public String allapp() {
    ActionContext.getContext().put("applist", appList);
    return "success";
}
\end{lstlisting}

This method processes student password modification requests within the MVC based system architecture. The controller layer first verifies the consistency between the new password and its confirmation to ensure input validity. It then invokes the business logic layer to execute the password update operation under controlled transactional conditions(See Listing \ref{code3} for Details). The system implements input validation, exception handling, and secure password storage to enhance data integrity and operational security. Based on the result returned by the business layer, the controller returns the corresponding success or failure view state, thereby reinforcing the separation of concerns and maintainability within the layered architecture\cite{4}\cite{rao2024data}.

\begin{lstlisting}[language=Java,caption={Modify Information},label={code3}]
public String change() {
    if (!npassword.equals(confirmPassword)) {
        return "changefail"; 
    }

    boolean success = studentService.changePassword(student.getId(), oldPassword, npassword);
    if (success) {
        return "changesuccess";
    } else {
        return "changefail";
    }
}
\end{lstlisting}

This method enables students to query their scheduled presentation arrangements within the system. The service layer first verifies the student’s identity to ensure secure and authorized access(See Listing \ref{code4} for Details). It then retrieves all scheduled activities associated with the student and stores the resulting data in the request context. Finally, the controller returns a success status to trigger the appropriate view rendering, thereby completing the full process from data retrieval to presentation. This design reinforces secure data access, preserves data integrity, and illustrates the separation of concerns within the layered architecture\cite{john2024leveraging}\cite{artar2024improving}.

\begin{lstlisting}[language=Java,caption={Inquire About Registration},label={code4}]
public String findMyArrangements() {
    Student currentStudent = studentService.findById(student.getId());
    if (currentStudent == null) {
        return "error";
    }
    List<Arrangement> arrangements = arrangementService.findMyArrangements(student.getId());
    ActionContext.getContext().put("list", arrangements);
    return "success";
}
\end{lstlisting}

This method implements the functionality that allows students to register for promotional events. The service layer first validates both the student’s credentials and the event information, and checks whether a registration record already exists. If no prior registration exists, it executes the registration operation and returns a success confirmation. If the student is already registered or an exception occurs, the method returns the corresponding error message. This design demonstrates comprehensive business logic control and incorporates robust exception handling mechanisms to ensure system reliability and data integrity(See Listing \ref{code5} for Details).

\begin{lstlisting}[language=Java,caption={Register for Event},label={code5}]
public String addSa() {
    try {
        Student s = studentService.findById(student.getId());
        Arrangement a = arrangementService.findById(arrangement.getId());
        if (arrangementService.isApplied(s, a)) {
            addActionError("Success!");
            return "error";
        }
        arrangementService.addSa(s,a);

        addActionMessage("Success!");
        return "success";
    } catch (Exception e) {
        addActionError("Failed:" + e.getMessage());
        return "error";
    }
}
\end{lstlisting}

This controller class handles requests related to presentation activities. It invokes the findAll() method in the service layer to retrieve the complete list of activities and stores the results in the arclist property(See Listing \ref{code6} for Details). Upon returning a  success  status, the corresponding view renders the data for display. This design demonstrates the loose coupling between the controller and the view within the MVC architecture, thereby enhancing maintainability and scalability\cite{calegari2019web}.

\begin{lstlisting}[language=Java,caption={Modify Presentation Information},label={code6}]
public class ArrangementAction {
    private List<Arrangement> arclist; 
    private Student student;


    public String findAll() {
        arclist = arrangementService.findAll();
        return "success";
    }

    public List<Arrangement> getArclist() {
        return arclist;
    }

    public void setArclist(List<Arrangement> arclist) {
        this.arclist = arclist;
    }
}
\end{lstlisting}

This service class provides functionality for querying presentation event details and utilizes HQL for data access. The findDetails method returns a single event entity after parameter validation, whereas findDetailWithCompanyInfo returns a DTO object that encapsulates event information together with associated company data through a join query. The code demonstrates business encapsulation and exception handling mechanisms at the service layer, thereby ensuring clear separation of concerns and robust data management(See Listing \ref{code7} for Details).

\begin{lstlisting}[language=Java,caption={Query Presentation Event Information},label={code7}]
@Service
@Transactional(readOnly = true)
public class ArrangementService {    
    @Autowired
    private ArrangementDAO arrangementDAO;
    
    public Arrangement findDetails(Integer arrangementId) {
        if (arrangementId == null || arrangementId <= 0) {
            throw new ValidationException("No Failed");
        }
        
        try {
            String hql = "FROM Arrangement WHERE id = :arrangementId";
            Map<String, Object> params = new HashMap<>();
            params.put("arrangementId", arrangementId);
            
            List<Arrangement> arrangements = arrangementDAO.findByHQL(hql, params);
            
            if (arrangements != null && !arrangements.isEmpty()) {
                return arrangements.get(0);
            }
            
            return null;
        } catch (DataAccessException e) {
            logger.error("ID: " + arrangementId, e);
            throw new ServiceException("Failed");
        } catch (Exception e) {
            logger.error("Query ID: " + arrangementId, e);
            throw new ServiceException("Failed");
        }
    }

    public ArrangementDetailDTO findDetailWithCompanyInfo(Integer arrangementId) {
        String hql = "SELECT new com.example.dto.ArrangementDetailDTO(" +
                    "a.id, a.title, a.description, a.arrangeTime, a.location, " +
                    "a.applyDeadline, a.maxParticipants, a.applyCount, " +
                    "c.id, c.name, c.industry, c.scale, c.description) " +
                    "FROM Arrangement a " +
                    "JOIN a.company c " +
                    "WHERE a.id = :arrangementId";        
        Map<String, Object> params = Collections.singletonMap("arrangementId", arrangementId);        
        List<ArrangementDetailDTO> result = arrangementDAO.findByHQL(hql, params);
        return result != null && !result.isEmpty() ? result.get(0) : null;
    }
}
\end{lstlisting}

This controller method implements the query and display functionality for seminar event details(See Listing \ref{code8} for Details). It first validates parameter integrity, then retrieves event detail data through the service layer and checks the current student's registration status. The method passes data to the view layer via property setters and finally returns the corresponding logical view name based on the operation result. This process demonstrates a complete business data preprocessing workflow within the MVC architecture.

\begin{lstlisting}[language=Java,caption={Query Seminar Registration},label={code8}]
public String findDetails() {
    try {
        if (arrangement == null || arrangement.getId() == null) {
            addActionError("No");
            return "detailFail";
        }

        Arrangement arrangementDetail = arrangementService.findDetails(arrangement.getId());

        if (arrangementDetail == null) {
            addActionError("No");
            return "detailFail";
        }

        setArrangementDetail(arrangementDetail);

        Student currentStudent = getCurrentStudent();
        if (currentStudent != null) {
            boolean hasApplied = arrangementService.checkStudentApplied(arrangement.getId(), currentStudent.getId());
            setHasApplied(hasApplied);
        }

        return "detailSuccess";
    } catch (Exception e) {
        addActionError("Failed");
        logger.error("Failed", e);
        return "error";
    }
}
\end{lstlisting}

This method implements unified search routing functionality by dynamically invoking different business services based on the search type parameter. When the search type corresponds to job postings, it calls the job search service; when it corresponds to event information, it calls the event search service(See Listing \ref{code9} for Details). The return value directs users to the corresponding results page, thereby embodying the fundamental principles of the Strategy pattern and enhancing flexibility and scalability in the system’s search mechanism.

\begin{lstlisting}[language=Java,caption={Searching for Event Information},label={code9}]
public String search() {
    if ("Recruit".equals(search.getSearch_type())) {
        lists = recruitService.searchJob(search);
        return "JobSuccess";
    } else if ("Application".equals(search.getSearch_type())) {
        lists = recruitService.searchApp(search);
        return "AppSuccess";
    } else {
        return "input";
    }
}
\end{lstlisting}

This method implements the query functionality for student resume records. The controller layer calls the service layer to retrieve the current student resume list, while the service layer queries the database using HQL and sets the results directly in the HttpServletRequest properties(See Listing \ref{code10} for Details). Although the code adopts a layered architecture, the service layer directly manipulates the request object, which introduces architectural coupling issues and reduces the separation of concerns between layers.

\begin{lstlisting}[language=Java,caption={View Resume Information},label={code10}]
public String findMy()/*Action*/

{
    studentService.findById(student);
    recruitService.findMy(student);
    return "success";

}

public void findMy(Student student) /*Service*/

{
    ActionContext ctx=ActionContext.getContext();
    request=(Map)ctx.get("request");
    re = (HttpServletRequest)ctx.get(ServletActionContext.HTTP_REQUEST);
    String hql = "SELECT * FROM resume WHERE student_id="+student.getId();
    List list = recruitDAO.findResume(hql);
    re.setAttribute("list",list);
}
\end{lstlisting}
\section{Evaluation}
\subsection{Test Objectives}
In the field of software engineering, as system architecture and functional complexity continue to grow, defects inevitably arise during the development process\cite{20}. These defects often lurk in subtle aspects such as code logic, module interfaces, or data processing, thereby posing potential threats to system stability, security, and ultimately the user experience. Therefore, software testing constitutes an indispensable and critical phase in the software development lifecycle. It functions not merely as an error detection process but serves as a rigorous engineering activity aimed at enhancing product quality. Through systematic methods, it involves designing test cases, executing procedures, and evaluating results to verify whether the software meets specified requirements and to uncover hidden errors. Comprehensive testing significantly reduces system risks, ensures reliable product delivery, and ultimately enhances user satisfaction\cite{15}.

\subsubsection{Testing Process}
This testing effort primarily employs black box testing methods, with the core objective of verifying whether software functionality meets predetermined requirement specifications without focusing on internal code implementation logic\cite{21}. The testing process simulates typical operational behaviors of real user roles (including students, enterprises, and administrators), designs test cases covering normal, abnormal, and boundary conditions, and inputs corresponding test data. By comparing the program’s actual output with expected results, the correctness, completeness, and user interaction compliance of each functional module are systematically validated. Any defects or abnormal behaviors discovered during testing are meticulously documented, including reproduction steps, actual results, and deviations from expectations. These findings enter the defect management system for tracking and management, thereby providing clear evidence for subsequent code debugging and version iterations\cite{16}.

\subsubsection{Student Side Testing}
In the student side functional testing, based on the Tomcat v9.0 environment, the system undergoes systematic functional validation using black box testing methods. Testing covers four core modules: identity authentication, personal information management, recruitment information processing, and participation in promotional events. By simulating user behavior, the correctness and stability of key processes—including login logic, data display and modification, resume submission, and event registration—are verified. Test results indicate that all functions operate as expected, with complete interaction workflows and sound data consistency. The system achieves its intended design objectives and meets the quality standards for deployment\cite{17}.

\subsubsection{Enterprise Side Testing}
Functional testing for the enterprise side occurs in a Tomcat v9.0 environment. The system validates core functionalities for enterprise users, including full lifecycle management of recruitment information, resume processing, application for promotional events, and corporate information maintenance. Test results confirm that, after logging in, enterprise users fully execute operations to publish, query, modify, and delete recruitment information, as well as view details and respond to the status of submitted resumes. Additionally, functions such as publishing and querying presentation information, displaying basic company details, and modifying passwords operate stably. Page transitions and data updates across all stages remain accurate, confirming that the enterprise side modules comply with business design specifications and deliver complete, reliable functionality\cite{18}.

\subsubsection{Administrator Console Testing}
Under the Tomcat v9.0 testing environment, the system performs systematic verification of the administrator console’s core functionalities. Testing demonstrates that administrators effectively execute review and scheduling operations for presentation applications, successfully managing application status transitions. Personal information management and password modification functions operate stably, with permission controls and data display proving accurate and reliable. Overall, the system administration module meets design expectations and supports efficient and secure management operations\cite{19}.
\section{CONCLUTION}
The design and implementation of this system establish an intelligent recruitment management platform based on a B/S architecture and the J2EE three tier framework. By integrating the SpringBoot framework with SSH components and a MySQL database, the system demonstrates robust stability, security, and maintainability. Test results confirm that the three core modules student interface, enterprise interface, and administrator interface achieve complete functional implementation. Business processes operate smoothly, and user interactions align with expectations, thereby validating the rationality of the system design and the feasibility of the technical solution. Although the system achieves its fundamental design objectives, opportunities for improvement remain in areas such as high concurrency processing, interface responsiveness optimization, and mobile device adaptation. Future enhancements can incorporate caching mechanisms and asynchronous processing to boost system performance, while expanding data analytics capabilities to elevate the intelligence of recruitment processes, thereby better serving the campus recruitment ecosystem\cite{parsons2014influences}.

\bibliographystyle{plain}  % 可选样式：plain, unsrt, ieeetr 等
\bibliography{references}  % 指定 .bib 文件名（无扩展名）

\end{document}